\documentstyle[12pt,axodraw,epsfig,prc,aps]{revtex}
\newcommand{\fslash}[1]{\mbox{$\!\not\!#1$}}

\begin{document} 

\draft
\title{$\pi - \pi$ Scattering Lengths and Chiral Condensate in NJL Model}

\author{Mei Huang$^{1}$, Pengfei Zhuang$^{2}$, Weiqin Chao$^{1,3}$ \\ 
{ \small $^1$ Institute of High Energy Physics, Chinese Sciences Academy, Beijing 100039, China\\
  $^2$ Physics Department, Tsinghua University, Beijing 100084, China\\
  $^3$ CCAST, Beijing 100080, China}}
\date{\today}

\maketitle

\vskip 1cm

\begin{abstract}
Within the general framework of  
SU(2) Nambu-Jona-Lasinio model beyond mean-field approximation, 
the S-wave $\pi - \pi$ scattering lengths $a_0^0$ and $a_0^2$ are calculated.
The numerical results are in agreement with that analyzed from recent experiment 
$\pi N \rightarrow \pi \pi N$ data by using Chew-Low-Goebel technique,
and the corresponding chiral condensate $<{\bar q}q>$ is
around $-(250 {\rm MeV})^3$, which is close to 
the ones obtained from lattice and sum rules. 
\end{abstract}

\newpage
\section{Introduction}
~~Spontaneous chiral symmetry breaking (S$\chi$SB) is one of the important 
characteristics of QCD.
The standard scenario assumes that the mechanism of S$\chi$SB is triggered 
by the formation 
of a large $<{\bar q}q>$ in the QCD vacuum.
And there is an alternative scenario regarding that S$\chi$SB
can be realized even with a small or vanishing $<{\bar q}q>$
in confining vector-like gauge
theories \cite{stern1} \cite{stern2}. 

$\pi - \pi$ scattering at threshold will provide a test to the 
two pictures
of S$\chi$SB. For a summary of current experimental information can refer 
Ref. \cite{exp1}. In the near future, there will
be several experiments dedicated to the precise determination of the
$\pi - \pi$ scattering lengths, including  
semileptonic $K_{e4}$ decays from E865 \cite{e8} and KLOE \cite{kloe},
and the measurement of the pionium lifetime (DIRAC) \cite{dirac}, which directly
determines the S-wave $\pi - \pi$ scattering length difference $a_0^0-a_0^2$.

On the theoretical side, the S-wave $\pi - \pi$ scattering lengths
$a_0^0$ and $a_0^2$
have been investigated in the standard scenario by using the standard chiral 
perturbation theory 
up to two-loops \cite{gasser} \cite{bijnens} and the Nambu-Jono-Lasinio (NJL) 
model to the leading order of $1/N_c$ expansion where $N_c$ is the number
of color freedoms \cite{NJL1} - \cite{schu}. 
Also $a_0^0$ and $a_0^2$ have been calculated in the alternative scenario 
\cite{stern4} by using the generalized
chiral perturbation theory formulated by Orsay group \cite{stern3}.

The standard scenario of S$\chi$SB is usually motivated
by the NJL model \cite{NJL}, in which S$\chi$SB
is associated with large quark condensate. 
The effect of explicit chiral symmetry breaking induced by a small current
quark mass $m_0$ which is important to discuss the process of
$\pi - \pi$ scattering at threshold. As we have analyzed in \cite{our}, 
only beyond mean-field approximation, can the effect of massive pions  
be seen clearly. Our interest in this paper is to study the
S-wave $\pi - \pi$ scattering lengths 
$a_0^0$ and $a_0^2$ in a self-consistent
SU(2) NJL model \cite{Ann} with explicit chiral symmetry breaking \cite{our}.

\section{SU(2) NJL Model Beyond Mean-field Approximation}
~~The two-flavor NJL model is defined through the Lagrangian density,
\begin{eqnarray}
\label{lagr}
{\cal L} = \bar{\psi}(i\gamma^{\mu}\partial_{\mu}-m_0)\psi + 
  G[(\bar{\psi}\psi)^2 + (\bar{\psi}i\gamma_5{\bf {\vec \tau}}\psi)^2 ],
\end{eqnarray}
where $G$ is the effective coupling constant with dimension ${\rm GeV}^{-2}$, 
$m_0$ the current quark mass,
and $\psi, \bar{\psi}$ quark fields with flavor, colour and spinor indices
suppressed, assuming isospin degeneracy of the 
$u$ and $d$ quarks.

We first briefly review the general scheme of the NJL model beyond mean-field 
approximation \cite{our}.
Including current quark mass explicitly, 
the quark self-energy $m$ expanded to $O(1/N_c)$ order can be expressed as
\begin{eqnarray}
\label{gap}
m = m_0 + m_H +\delta m, 
\end{eqnarray}
where $m_H$ and $\delta m$ are the leading $O(1)$ and subleading  
$O(1/N_c)$ contributions shown in Fig. 1. The solid 
lines in Fig. 1
indicate quark propagator $S(p)=1/({\fslash p}-m)$ with full $m$.
The quark condensate $<{\bar q}q>$ is a one-loop quark integral 
\begin{eqnarray}
\label{cond}
<{\bar q}q>=\frac{1}{N_f}<{\bar \psi} \psi>
           =-4iN_c m \int\frac{d^4p}{(2\pi)^4}\frac{1}{p^2-m^2}.
\end{eqnarray}
The corresponding meson propagator $D_{M}(k)$ (M means 
$\pi$ or $\sigma$) has the form
\begin{eqnarray}
\label{prog}
-{\rm i} D_{M}(k) & = & \frac{2iG}{1-2G \Pi_{M}(k)}, \nonumber \\
\Pi_{M}(k) & = & \Pi_{M}^{(RPA)}(k)+\delta\Pi_{M}^{(b)}(k)+
             \delta\Pi_{M}^{(c)}(k)+\delta\Pi_{M}^{(d)}(k),
\end{eqnarray}
where $\Pi_{M}$ is the meson polarization function, which includes
the leading
order $\Pi_{M}^{(RPA)}$ and subleading order $\delta\Pi_{M}^{(b,c,d)}$, 
shown in Fig. 1. The above constituent quark mass and the meson propagator,
namely the Eqs. ~(\ref{gap}) and (\ref{prog}), or the Feynman diagrams
in Fig. 1,~ form a self-consistent description of the SU(2) NJL model to the
subleading order of $1/N_c$ expansion, which is different from the earlier
calculations \cite{NJL1} - \cite{schu} where only the mean-field quark
mass $m_H$ and the Random-Phase-Approximation (RPA) meson polarization function 
$\Pi_{M}^{(RPA)}$ are considered. 

The meson mass $m_{M}$ satisfies the total meson propagator's pole condition
\begin{eqnarray}
\label{pole}
1-2G\Pi_{M}(k^2=m_{M}^2)=0,
\end{eqnarray}
and the meson-quark coupling constant $g_{M qq}$ is determined by the residue at the pole
\begin{eqnarray}
\label{couple}
g_{M qq}^{-2} = (\partial \Pi_{M}(k)/\partial k^2)^{-1}|_{k^2=m_{M}^2}.
\end{eqnarray}
Another important quantity in the meson sector is the pion decay constant $f_{\pi}$ 
which 
generally satisfies 
\begin{eqnarray}
\label{onshell}
\frac{m_{\pi}^2f_{\pi}}{g_{\pi qq}} = \frac{m_0}{2G}.
\end{eqnarray}
In the chiral limit, $f_{\pi}$ satisfies the Goldberger-Treiman relation 
$f_{\pi}(k)g_{\pi qq}(k)=m$.

\begin{center}
\begin{picture}(500,100)(0,0)
\SetWidth{2}
\CArc(120,50)(15,0,360)
\Vertex(120,35){2.5}
\Text(120,19)[]{$m_H$}
\CArc(200,50)(15,0,360)
\DashCArc(200,70)(15,320,220){4}
\Vertex(200,35){2.5}
\Vertex(189,61){2.5}
\Vertex(210,61){2.5}
\Text(200,19)[]{$\delta m$}
\Text(200,75)[]{$\pi$,$\sigma$}
\end{picture}
\begin{picture}(500,85)(0,0)
\SetWidth{2}
\CArc(105,50)(15,0,360)
\Vertex(90,50){2.5}
\Vertex(120,50){2.5}
\Text(110,15)[]{$\Pi^{(RPA)}_{M}(k)$}
\CArc(180,50)(15,0,360)
\DashLine(180,65)(180,35){4}
\Vertex(165,50){2.5}
\Vertex(180,65){2.5}
\Vertex(180,35){2.5}
\Vertex(195,50){2.5}
\Text(180,15)[]{$\delta \Pi^{(b)}_{M}(k)$}
\Text(180,50)[]{$\pi$,$\sigma$}
\CArc(260,50)(15,0,360)
\DashCArc(260,70)(15,320,220){4}
\Vertex(249,60){2.5}
\Vertex(248,41){2.5}
\Vertex(271,60){2.5}
\Vertex(272,41){2.5}
\Text(260,75)[]{$\pi$,$\sigma$}
\Text(260,15)[]{$\delta \Pi^{(c)}_{M}(k)$}
\SetWidth{2}
\CArc(340,50)(10,0,360)
\CArc(400,50)(10,0,360)
\Vertex(330,50){2.5}
\Vertex(340,60){2.5}
\Vertex(340,40){2.5}
\Vertex(410,50){2.5}
\Vertex(400,60){2.5}
\Vertex(400,40){2.5}
\Text(370,65)[]{$\pi$}
\Text(370,35)[]{$\sigma$}
\Text(370,15)[]{$\delta \Pi^{(d)}_{M}(k)$}
\DashCurve{(340,60)(360,70)(380,70)(400,60)}{3.2}
\DashCurve{(340,40)(360,30)(380,30)(400,40)}{3.2}
\end{picture}
\end{center}
\begin{figure}[b]
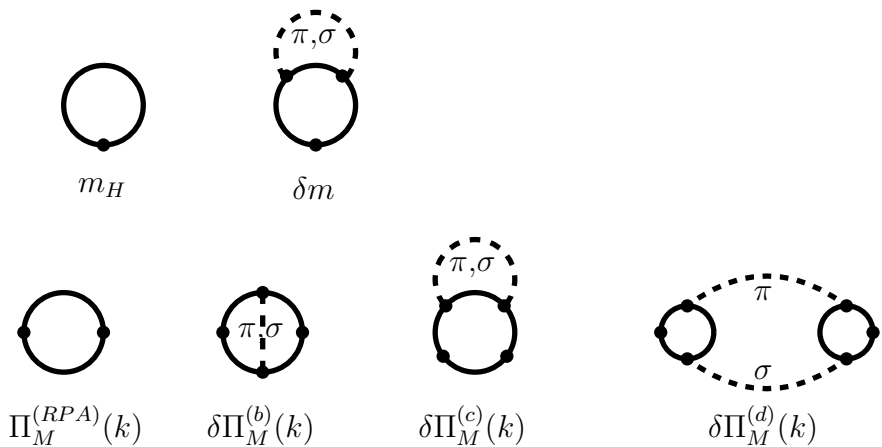

\begin{center}
\caption{Feynman digrams for leading and subleading order quark self-energy 
$m_H$ and $\delta m$, and meson polarization functions $\Pi^{(RPA)}_{M}$ and 
$\delta \Pi^{(b,c,d)}_{M}$. The solid and dashed lines indicate the quark 
internal meson propagators respectively.}
\label{kernelb}
\end{center}
\end{figure}

\section{$\pi - \pi$ scattering at threshold}

Now we turn to the calculation of the
S-wave $\pi - \pi$ scattering length $a_0^0$ and $a_0^2$. 
The invariant scattering amplitude can be generally written as \cite{sca}
\begin{eqnarray}
T_{ab,cd}=A(s,t,u)\delta_{ab} \delta_{cd}
          + B(s,t,u)\delta_{ac} \delta_{bd} + C(s,t,u) \delta_{ad} \delta_{bc},
\end{eqnarray}
where $a, b$ and $c, d$ are the isospin labels of the initial and final states 
respectively, and $s, t$ and $u$ are the Mandelstam variables, $s=(p_a+p_b)^2$,
$t=(p_a-p_c)^2$, $u=(p_a-p_d)^2$, 
where $p_a \sim p_d$ are initial and final isospin momentum.
Using perfect crossing symmetry, one can 
project out isospin amplitudes
\begin{eqnarray}
A_0=3A+B+C, \ \ \ A_1=B-C, \ \ \ A_2=B+C.
\end{eqnarray}
In the limit of scattering at threshold, ${\sqrt s}=2m_{\pi}$, t=u=0, the S-wave 
scattering lengths $a_0^{I}$ (given in units of $m_{\pi}^{-1})$ are:
\begin{eqnarray}
a_0^{I}=\frac{1}{32 \pi}A_I(s=4m_{\pi}^2, t=0, u=0), \ \ I=0,2.
\end{eqnarray}

We will calculate the scattering length to the 
lowest order in $1/N_c$ expansion of the process $\pi \pi \rightarrow \pi \pi$
in the NJL model with quark mass $m$ and meson polarization function $\Pi_M$ 
including the subleading order of $1/N_c$ expansion. The Feynman diagrams 
include box and $\sigma-$
exchange diagrams, and are the same as those  
in \cite{schu}, but in\cite{schu} quark mass $m_H$ in mean-field approximation and 
meson polarization function in RPA $\Pi_M^{(RPA)}$. 
 
To keep the lowest order of the $\sigma$ exchange diagrams, 
the internal $\sigma$ propagator  
should be in leading $O(1/N_c)$ order,
which has the same form as that in RPA $\Pi_M^{(RPA)}$ \cite{our}.
So what we need to do is using the expression formulae given in \cite{schu} 
directly,
and replacing the quantities in the mean-field approximation with those beyond
mean-field approximation, i.e.,
the external pion propagator becomes the total propagator Eq. (\ref{prog}),
the pole for quark propagator becomes the total quark self-energy $m$ defined in 
Eq. (\ref{gap}), and the coupling constant 
$g_{\pi qq}$ now is the one expressed in Eq. (\ref{couple}). 

\section{Numerical Results}
For the numerical calculations, we adopt the external
momentum expansion method as discussed in detail in \cite{our}.
 We introduce a quark momentum cut-off $\Lambda_f$ in 
Pauli-Villars regularization 
and a meson momentum cut-off $\Lambda_b$ in covariant regularization 
for the divergent momentum integrals. Using only the two experimental observables 
$m_{\pi}=139 {\rm MeV}$ and
$f_{\pi}=92.4 {\rm MeV}$ and the reasonable quark condensate $<{\bar q}q >^{1/3}$ 
in the empirical range of
 $-300 {\rm MeV} \sim -200 {\rm MeV}$ as input,
we can not give fixed values of the four parameters in the model, namely the 
current quark mass $m_0$,
coupling constant $G$, and the two momentum cuts $\Lambda_f$ and $\Lambda_b$. 
So we should introduce
one more free parameter $z=\Lambda_b / \Lambda_f$, 
which characterizes the meson cloud 
contributions. Especially,
in the limit of $z=0$, the model goes back to the
mean-field approximation automatically.

As we discussed in \cite{our}, for each $z$, quark condensate 
$-<{\bar q} q>^{1/3}$ has a minimum $(-<{\bar q} q>^{1/3})_{min}$,
around which there is a region quark
condensate changes slowly with quark mass $m$, and in this region of $m$,
$m_0$ and 
$\Lambda_f$ are in plateaus too. In Table. 1, we list the values of $m$, $m_0$ and
$\Lambda_f$ corresponding to $(-<{\bar q} q>^{1/3})_{min}$ (in units of GeV).
\vskip 0.2cm
\begin{center}
\begin{picture}(320,100)(0,0)
\SetWidth{1}
\Line(0,0)(320,0)
\Line(0,80)(320,80)
\Line(0,100)(320,100)
\Text(100,90)[]{$z=0$}
\Text(40,70)[]{$-<{\bar q} q>^{1/3}$}
\Text(40,50)[]{$m$}
\Text(40,30)[]{$m_0$}
\Text(40,10)[]{$\Lambda_f$}
\Text(280,70)[]{0.256}
\Text(220,70)[]{0.237}
\Text(160,70)[]{0.216}
\Text(160,90)[]{$z=0.5$}
\Text(220,90)[]{$z=1$}
\Text(280,90)[]{$z=1.5$}
\Text(100,70)[]{0.209}
\Text(280,50)[]{0.34}
\Text(220,50)[]{0.36}
\Text(160,50)[]{0.47}
\Text(100,50)[]{0.49}
\Text(280,30)[]{0.0067}
\Text(220,30)[]{0.0075}
\Text(160,30)[]{0.0086}
\Text(100,30)[]{0.0089}
\Text(280,10)[]{0.82}
\Text(220,10)[]{0.73}
\Text(160,10)[]{0.63}
\Text(100,10)[]{0.615}
\end{picture}
\end{center}
\vspace*{-0.2truecm}
{\small Table. 1. The values of $m$, $m_0$ and
$\Lambda_f$ corresponding to $(-<{\bar q} q>^{1/3})_{min}$ (in units of GeV) for 
different $z=\Lambda_b / \Lambda_b$.}

For each $z$, with changing $<{\bar q}q>$, we get a series of 
$m$ and $g_{\pi qq}$, from which we can calculate the S-wave $\pi - \pi $ scattering 
length $a_0^0$ and $a_0^2$. In the quark condensate plateau region around 
$-<{\bar q} q>^{1/3}_{min}$, there is also a plateau for $a_0^0$ and $a_0^2$
respectively.  
We choose the values of $a_0^0$ and $a_0^0-a_0^2$ corresponding to 
$(-<{\bar q} q>^{1/3})_{min}$
to compare with those analyzed from experimental data and 
predicted by other 
theories, which are listed in Table. 2, where all
the values are in units of $m_{\pi}^{-1}$.
In this table, the experimental values of $a_0^0$ and $a_0^2$  
without star were obtained in 1979 in
a comprehensive phase shift analysis of peripheral $\pi N \rightarrow \pi \pi N$
reactions and $K_{e4}$ decays \cite{exp2}; 
those marked with star
were from the Chew-Low-Goebel analyses of the recent CHAOS 
data $\pi^{-}p \rightarrow \pi^{+}\pi^{-}n$, and  LAMPF E1179 data
$\pi^{+}p \rightarrow \pi^{+}\pi^{0}p$ \cite{exp3}.  
And the experimental value of $a_0^0 - a_0^2$
was obtained from the universal curve in \cite{bijnens}.
\vskip 0.2cm
\begin{center}
\begin{picture}(460,120)(0,0)
\SetWidth{1}
\Line(0,120)(460,120)
\Line(0,100)(460,100)
\Line(0,0)(460,0)
\Line(0,60)(460,60)
\Line(0,20)(460,20)
\Text(20,10)[]{$a_0^0-a_0^2$}
\Text(20,40)[]{$a_0^2$}
\Text(20,80)[]{$a_0^0$}
\Text(300,110)[]{$exp$}
\Text(300,90)[]{0.26$\pm$0.05}
\Text(300,50)[]{-0.028$\pm$0.012}
\Text(300,70)[]{* 0.206$\pm$0.013}
\Text(300,30)[]{* -0.055$\pm$0.021}
\Text(300,10)[]{0.29$\pm$0.04}
\Text(370,110)[]{$a$}
\Text(370,80)[]{0.217}
\Text(370,40)[]{-0.0413}
\Text(370,10)[]{0.258}
\Text(430,110)[]{$b$}
\Text(430,80)[]{0.263}
\Text(430,40)[]{-0.027}
\Text(430,10)[]{0.29}
\Text(70,110)[]{NJL($z=0$,}
\Text(70,80)[]{0.16}
\Text(70,40)[]{-0.044}
\Text(70,10)[]{0.20}
\Text(120,110)[]{$z=0.5$,}
\Text(120,80)[]{0.17}
\Text(120,40)[]{-0.046}
\Text(120,10)[]{0.216}
\Text(170,110)[]{$z=1$,}
\Text(170,80)[]{0.196}
\Text(170,40)[]{-0.052}
\Text(170,10)[]{0.248}
\Text(220,110)[]{$z=1.5$)}
\Text(220,80)[]{0.221}
\Text(220,40)[]{-0.057}
\Text(220,10)[]{0.278}
\end{picture}
\end{center}
\vspace*{-0.2truecm}
{\small Table. 2. Scattering lengths predicted by NJL model in different $z$ 
compare with the experimental values $(exp)$ and those calculated from
chiral perturbation theory $(a)$ \cite{bijnens} and generalized chiral 
perturbation theory $(b)$ \cite{stern4}
up two-loop.}

From Table. 2 we can see that, $a_0^0$ and $a_0^2$ calculated from
NJL model at a certain $z$ between $1 \sim 1.5$ are in agreement with the recent 
experimental values marked with star, and the scattering lengths at 
$z=0$ are the same as the Weinberg values 
$a_0=0.16$ and $a_2=-0.044$ \cite{weinberg}. Comparing 
with those at $z=0$, the scattering lengths at $1<z<1.5$ 
are all modified by $20 \sim 40 \% $.

In the region of $z$ between
$1 \sim 1.5$ in Table. 1, the quark mass $m$ is about 333 {\rm MeV}, and 
current quark mass $m_0$ is
in the empirical region $ 5 {\rm MeV} \sim 7 {\rm MeV}$, and
the quark condensate $-< {\bar q} q >$ is
between $(237 {\rm MeV})^3 \sim (256 {\rm MeV})^3$, which 
is in agreement with that $(245 \pm 4 \pm 9 \pm 7 {\rm MeV})^3$
from the lattice simulation \cite{lattice} 
and that $(242 \pm 9 {\rm MeV})^3$ from
the sum rule calculation (see \cite{srule} and references therein).

It is can be seen from Table. 2, that the recent experimental results 
support the standard chiral mechanism, i.e., S$\chi$SB is triggered by the 
formation of large quark condensate in QCD vacuum. 
While the scattering lengths predicted by two-loop 
generalized chiral perturbation theory, in which the quark condensate 
$<{\bar q}q>$ 
is around $(-100 {\rm MeV})^3$, are in agreement with 
the old experimental values without star. 
Anyway, which S$\chi$SB mechanism is right, should be answered 
after the analyses of
the new experimental data in the near future.

If the elastic pion-pion scattering experimental data in the near future 
support the mechanism of large quark condensate, 
the free parameter $z$ in the NJL model introduced to 
characterize the mesonic contributions
can be fixed by the scattering lengths.  
Then one can determine
the four parameters $m_0$, ${\rm G}$, $\Lambda_f$ and $\Lambda_b$.
From the above analyses, it seems that $z$ is between $1 \sim 1.5$ 
for the NJL model.

\section*{Acknowledgement}
~~The authors would like to thank Dr. Bing-Song Zou for his kind help during
the work. This work is supported by NNSF of China
(Nos. 19677102 and 19845001).

\end{document}